\documentclass[twocolumn,preprintnumbers,amsmath,amssymb]{revtex4} %RevTex 4 - NO*PACS
\usepackage{graphicx}
\usepackage{graphicx,color}% Include figure files
\usepackage{dcolumn}% Align table columns on decimal point
\usepackage{bm}% bold math
\begin{document}

%\preprint{APS/123-QED}
\preprint{AIP/123-QED}

\newcommand {\rsq}[1]{\langle R^2 (#1)\rangle}

%\title{Manuscript Title:\\with Forced Linebreak}% Force line breaks with \\
\title{Conformational statistics of randomly-branching double-folded ring polymers}

\author{Angelo Rosa}
% \altaffiliation[Also at ]{Physics Department, XYZ University.}%Lines break automatically or can be forced with \\
%\author{Second Author}%
\email{anrosa@sissa.it}
\affiliation{
%Authors' institution and/or address\\
%This line break forced with \textbackslash\textbackslash
Sissa (Scuola Internazionale Superiore di Studi Avanzati), Via Bonomea 265, 34136 Trieste, Italy
}

\author{Ralf Everaers}
\email{ralf.everaers@ens-lyon.fr}
\affiliation{
Universit\'e de Lyon, ENS de Lyon, UCBL, CNRS, Laboratoire de Physique and Centre Blaise Pascal, Lyon, France
}

\date{\today}% It is always \today, today,
             %  but any date may be explicitly specified

%%%
\begin{abstract}
The conformations of topologically constrained double-folded ring polymers can be described as wrappings of randomly branched primitive trees. 
We extend previous work on the tree statistics under different (solvent) conditions to explore the conformational statistics of double-folded rings in the limit of tight wrapping.
In particular, we relate the exponents characterizing the ring statistics to those describing the primitive trees and discuss the distribution functions $p(\vec r | \ell)$ and $p(L | \ell)$ for the spatial distance, $\vec r$, and tree contour distance, $L$, between monomers as a function of their ring contour distance, $\ell$.
\end{abstract}
%%%

% PACS, the Physics and Astronomy
                             % Classification Scheme.
%\keywords{Suggested keywords}%Use showkeys class option if keyword
                              %display desired
\maketitle

%%%
\section{Introduction}\label{sec:Intro}
%%%

\begin{figure}[!t]
%\begin{center}
\includegraphics[width=\columnwidth]{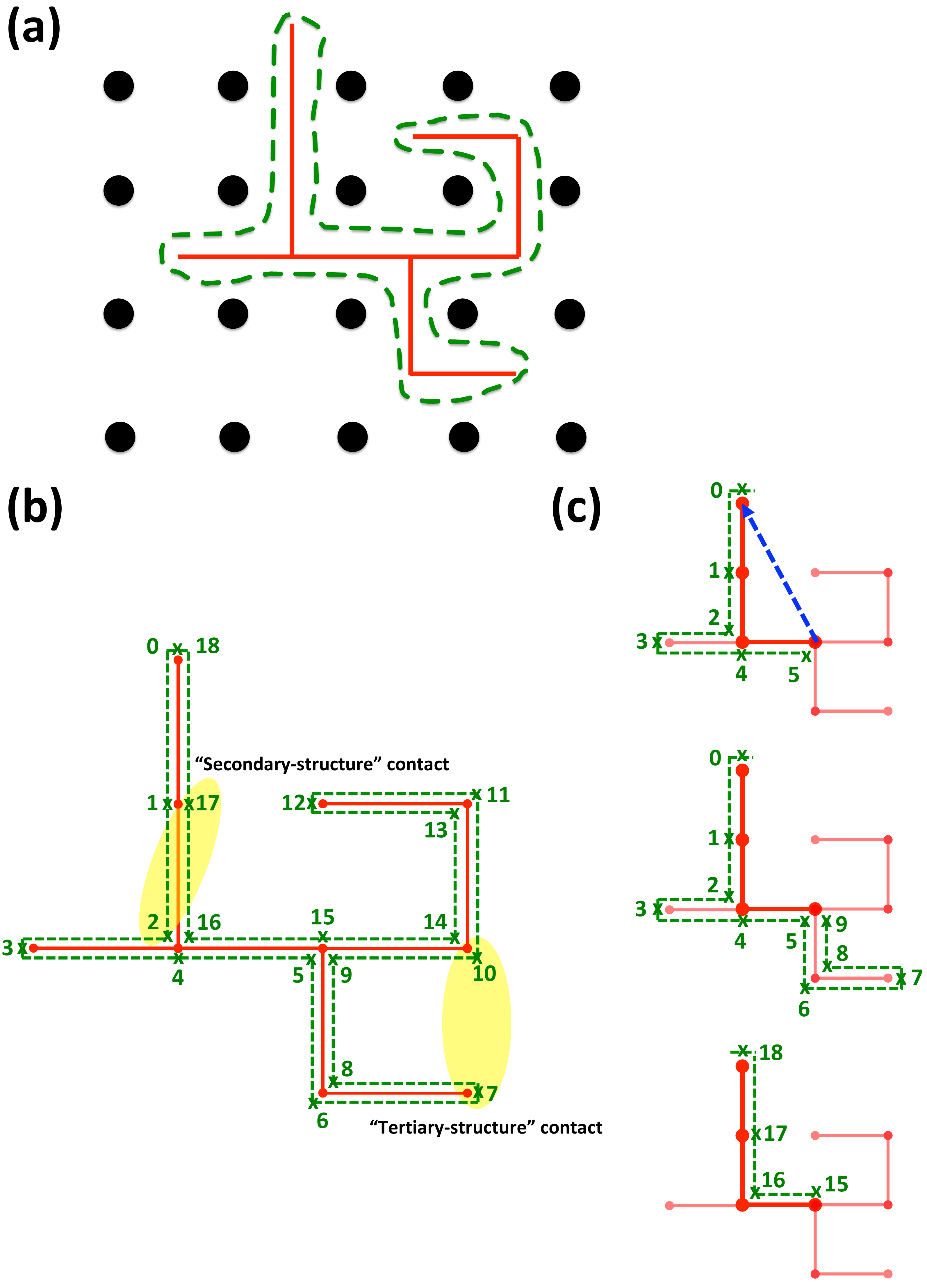}
%\end{center}
\caption{
\label{fig:DoubleFolding}
(a)
Ring polymers (green dashed line) in an array of fixed obstacles (and, in concentrated solutions and melt) adopt spatial conformations whose primitive paths (red solid line) can be mapped to randomly branched polymers with annealed connectivity.
(b)
Double-folded ring polymer (green dashed line) from a branched polymer (red solid line) of $N=9$ bonds.
The numbers indicate the contour distance of the ring (in units of elementary bonds) measured from one arbitrarily chosen end of the branched polymer.
The pairs of node $(2,17)$ and $(7,10)$ are classified as ``secondary-'' and ``tertiary-structure'' contacts, respectively.
(c)
Ring contour distances $\ell=5,9,3$ (dashed green lines, top to bottom) compatible with the tree contour distance $L=3$ (thick red line) and spatial distance $|\vec r|$ (blue arrow).
}
\end{figure}

Topologically constrained ring polymers often adopt double folded configurations characterized by a randomly branched primitive tree~\cite{KhokhlovNechaev85,RubinsteinPRL1986,RubinsteinPRL1994,GrosbergSoftMatter2014,RosaEveraersPRL2014,SmrekGrosberg2015,Michieletto-SoftMatter2016} (Figure~\ref{fig:DoubleFolding}(a)).
In analogy to protein or RNA structures~\cite{alberts}, such rings can be discussed in terms of a primary, a secondary, and a tertiary structure.
The primary structure is simply defined through the  connectivity of the ring monomers.
The secondary structure arises from the double folding. 
Its characterisation comprises information on the mapping of the ring onto the graph as well as on the connectivity of the graph representing the primitive tree.
Important characteristics are the contour distances, $L$, between tree nodes or the weight, $N_{br}$, of branches separated from the tree by severing a link. 
Finally, the tertiary structure defines the embedding of the rings and trees into (three dimensional) space. Relevant observables are the spatial distance, $\vec r$, between ring monomers or tree nodes, or the overall gyration radius, $R_g$.

As customary in polymer physics~\cite{DoiEdwards,RubinsteinColby}, the tree behavior can be analyzed in terms of a small set of exponents describing how expectation values for these observables vary with the weight, $N$, of the trees or the contour distance, $L$, between nodes.
Flory theory~\cite{IsaacsonLubensky,DaoudJoanny1981,GutinGrosberg93,GrosbergSoftMatter2014,RosaEveraers-SoftMatter2017} provides a useful framework for discussing the {\it average} behavior
of a wide range of interacting tree systems beyond the small number of available exact results~\cite{ZimmStockmayer49,DeGennes1968,ParisiSourlasPRL1981}. Furthermore, we have recently shown that the non-Gaussian distribution functions for tree observables are often of the Redner-des Cloizeaux (RdC) form
and characterized by a small set of additional exponents, which can often be related to each other and the standard trees exponents~\cite{RosaEveraers-PRE2017}.

Here we take a step back to the original polymer problem and consider rings, which are tightly wrapped around trees (Fig.~\ref{fig:DoubleFolding}(b)).
``Navigating'' on the tree along a wrapped ring mixes the two concepts invoked above for characterizing trees: 
(i) the branching statistics controls how fast the mean contour distance on the tree grows with the contour distance between monomers along the ring and
(ii) spatial distances depend on the conformations of paths on the tree in the embedding space. 
We consider ensembles of trees as described in our earlier works~\cite{RosaEveraers-JPA2016,RosaEveraers-JCP2016,RosaEveraers-PRE2017}, namely: single self-avoiding trees in three dimensions and melts of trees in two and three dimensions.
As a reference, we also consider the case of double-folded rings on conformations of {\it ideal} trees, {\it i.e.} without volume interactions.
Specifically, we tightly wrap ring polymers around tree conformations we generated in our previous Monte Carlo simulations and explore the relation between the tree and the ring exponents.

The paper is organized as follows:
In Section~\ref{sec:TheoBackgr},
we define the tight wrapping of a tree by a ring and briefly review definitions and the theoretical background.
We provide no explicit Methods section as there are no particular algorithmic difficulties associated to the wrapping process or the data analysis. 
The reader will instead find a concise introduction to the Monte Carlo methods used for generating lattice tree conformations and to the computational procedures for estimating scaling exponents and more numerical details in the Appendices at the end of the paper.
We present and discuss results in Section~\ref{sec:Results} and conclude in Section~\ref{sec:Concls}.

%%%
\section{Definitions and background}\label{sec:TheoBackgr}
%%%

%%%
\subsection{Topologically constrained ring polymers}
%%%

The transient folding of ring polymers subject to topological constraints (Fig.~\ref{fig:DoubleFolding}(a)) can be understood by using the analogy to randomly branched structures with annealed connectivity~\cite{KhokhlovNechaev85,RubinsteinPRL1986,RubinsteinPRL1994,kapnistos2008,GrosbergSoftMatter2014,RosaEveraersPRL2014,RosaEveraers-JPA2016,RosaEveraers-JCP2016}.
A typical example is the one of a single ring in an array of fixed obstacles~\cite{KhokhlovNechaev85,RubinsteinPRL1986,RubinsteinPRL1994} which represents a theoretical model for the problem of a ring moving through a gel~\cite{michieletto2017ring}.
Furthermore, several authors have suggested that the same analogy may be applied to describe spatial conformations of unknotted and untangled ring polymers in melts~\cite{CatesDeutsch,KhokhlovNechaev85,RubinsteinPRL1994,RubinsteinMacromolecules2016} and chromosomes in eukaryotes~\cite{grosbergEPL1993,RosaPLOS2008,Vettorel2009,MirnyRev2011}.
Although it remains a non-trivial question, if~\cite{CatesDeutsch} or to which extent~\cite{RubinsteinPRL1994,LangMacromol2013,RubinsteinMacromolecules2016,SmrekGrosbergACSMacroLett2016} the link to the melts can be pushed forward, we have shown~\cite{RosaEveraersPRL2014}  that it provides at least an excellent approximation.

In this work,
we consider double-folded ring polymers on lattice branched structures taken from different ensembles:
$3d$ trees in good solvent conditions {\it i.e} with purely repulsive interactions between tree nodes,
and
$2d$ and $3d$ melt of trees which are relevant to the problem of untangled polymers in dense solution.
For comparison, we also study single ring conformations whose branched primitive paths are generated under ideal conditions {\it i.e.} without excluded volume effects between tree nodes.

%%%
\subsection{Rings, trees, notation and units}
%%%

We measure energy in units of $k_BT$ and length in units of the Kuhn length, $l_K$, of the ring polymer.
There are two oppositely oriented ring segments associated to each tree segment.
As a consequence, there are $2N$ ring segments for a tree composed of $N$ segments. 
This corresponds to $N+1$ nodes for the tree and $2N+1$ nodes for the ring.
To see that it is possible to characterize the primitive trees and the rings by the same Kuhn length,
consider the limit of a small double-folded ring of contour length $\ell_{ring} \equiv 2N$ corresponding to an unbranched primitive chain.
The gyration radius of the latter is given by $R_g^2=l_K (\ell_{ring}/2)/6 = l_K \ell_{ring}/12$, corresponding to a closed random walk~\cite{RosaEveraersPRL2014}. 

We use the letters $N$ and $N_{br}$ to denote the number of segments of a tree or a branch, respectively.
The symbols $L$ and $\ell$ are reserved for contour lengths on the tree and the ring, respectively.
Spatial distances are denoted by the letters $R$ and $r$.
Examples are the tree gyration radius, $R_g$ and spatial distances between nodes, $\vec r_{ij}$.

A wrapping introduces an additional metric on the embedded graph, {\it i.e.} two nodes $I$ and $J$ can be characterized by their spatial distance,
$\vec r_{IJ} = \vec r_I - \vec r_J$, their contour distance, $L_{IJ}$, on the tree, and their contour distance, $L_{IJ} \le \ell_{IJ} \le N$, on the ring.
In the following, we substitute the arbitrary labelling $0 \le I,J \le N$ of the tree nodes by the (primary structure) ring labels $0 \le i,j \le 2N$ such that $\ell_{ij} \equiv \mbox{mod}(|i-j|,N)$.
Furthermore, we distinguish between secondary structure contacts on the tree and tertiary structure contacts occurring in a particular tree conformation in the embedding space.
Figure~\ref{fig:DoubleFolding}(c) illustrates $|\vec r_{ij}|$ (blue arrow), $L_{ij}$ (thick red line) and the corresponding $\ell_{ij}$'s (dashed green lines) for an example.

%%%
\subsection{Randomly branching trees}
%%%

A small set of exponents describes how expectation values for observables characterising tree connectivities and conformations vary with the weight, $N$, of the trees or the contour distance, $L$, between nodes:
\begin{eqnarray}
\langle N_{br}(N) \rangle &\sim& N^{\epsilon} \label{eq:epsilon}\\
\langle L(N) \rangle &\sim& N^\rho \label{eq:rho}\\
\langle R^2(L) \rangle &\sim& L^{2\nu_{\rm path}} \label{eq:nu_path}\\
\langle R_g^2(N) \rangle &\sim& N^{2\nu} \label{eq:nu}
\end{eqnarray}
Here, $\langle N_{br}(N)\rangle$ denotes the average branch weight;
$\langle L(N) \rangle$ the average contour distance or length of paths on the tree;
$\langle R^2(L) \rangle$ the mean-square spatial distance between nodes with fixed contour distance; and
$\langle R_g^2(N) \rangle$ the mean-square gyration radius of the trees.
By construction, $\nu = \rho \, \nu_{\rm path}$, and the relation $\epsilon=\rho$ is expected to hold in general~\cite{MadrasJPhysA1992}.
For ideal, non-interacting trees $\rho=\epsilon=\nu_{path}=1/2$ and $\nu=1/4$~\cite{ZimmStockmayer49,RosaEveraers-JPA2016}.
For interacting systems, the only exactly known exponent is $\nu=1/2$ for self-avoiding trees in $d=3$ dimensions~\cite{ParisiSourlasPRL1981}.
Otherwise, Flory theory~\cite{IsaacsonLubensky,DaoudJoanny1981,GutinGrosberg93,GrosbergSoftMatter2014,RosaEveraers-SoftMatter2017} provides a useful, although approximate, framework for discussing the {\it average} behavior, Eqs.~(\ref{eq:epsilon}) to~(\ref{eq:nu}), of a wide range of interacting tree systems beyond the small number of available exact results~\cite{ZimmStockmayer49,DeGennes1968,ParisiSourlasPRL1981}. 

The non-Gaussian nature of distribution functions for the above mentioned observables can be investigated by a combination of computer simulations and scaling arguments~\cite{RosaEveraers-PRE2017}. 
The branching statistics is in good agreement with a generalized Kramers form,
\begin{equation}\label{eq:branching probability}
p_{br}(N_{br}) \approx \left( \frac {N_{br} (N-N_{br})}{N} \right)^{-(2-\epsilon)} \stackrel{N_{br}\ll N}{\rightarrow}  N_{br}^{-(2-\epsilon)}\ .
\end{equation}
The end-to-end vector distributions, $p_N(\vec r|L) \equiv P_N(\left| \vec r \right| | \, L) \, / \, \frac{2\pi^{d/2}\left| {\vec r} \right|^{d-1}}{\Gamma(d/2)} $, for paths of length $L$ on trees of mass $N$ can be approximated as
\begin{equation}\label{eq:RdC-Ltree-E2E}
p_N(\vec r | L) = \frac1{\rsq{L}_N^{d/2}}\  q\left(\frac{\vec r}{ \sqrt{\rsq{L}_N}}\right) \, .
\end{equation}
Here, $\Gamma(z)$ denotes the Euler's gamma function, ${\vec x} = {\vec r} / \sqrt{\rsq{L}_N}$ is the scaled distance and
\begin{equation}
q(\vec x) = C \, |\vec x|^{\theta_{path}}\  \exp \left( -(K |\vec x|)^{t_{path}} \right)
\label{eq:q_RdC_path}
\end{equation}
is of the Redner-des Cloizeaux (RdC)~\cite{Redner1980,DesCloizeauxBook} form. 
The shape of the rescaled distributions, and hence the characteristic exponents $\theta_{path}$ and $t_{path}$ controlling the small and large distance behaviors respectively, depend on the universality class.
While $\theta_{path}$ is an independent exponent, $t_{path}$ can be estimated using the Fisher-Pincus relation~\cite{FisherSAWShape1966,PincusBlob1976}
\begin{equation}\label{eq:tPath}
t_{path} = \frac{1}{1-\nu_{path}} \, .
\end{equation}
The constants
\begin{eqnarray}
C &=& t \, \frac{\Gamma(1+\frac d2)\Gamma^{\frac{d+\theta}2}(\frac{2+d+\theta}t)}{d\,\pi^{d/2}\,\Gamma^{\frac{2+d+\theta}2}(\frac{d+\theta}t)}
\label{eq:RdC_C}\\
K^2 &=& \frac{\Gamma(\frac{2+d+\theta}t)}{\Gamma(\frac{d+\theta}t)} \ 
\label{eq:RdC_K}
\end{eqnarray}
are determined by the conditions (1) that the distribution is normalized ($\int q(\vec x) d\vec x \equiv 1 $) and (2) that the second moment was chosen as the scaling length ($\int |x|^2 q(\vec x) d\vec x \equiv 1 $).

%%%
\subsection{Tight wrapping of ring polymers}
%%%

The wrapping procedure for trees with a maximal functionality of $f=3$~\cite{RosaEveraersPRL2014} is schematically described in Fig.~\ref{fig:DoubleFolding}(b).
We start from a randomly chosen tip (numbered as ``$0$'' in the figure) and construct the ring by adding new bonds consecutively following the branched structure.
At a branching point (say, the one marked by the number ``2'') we choose randomly between one of the two possible directions and continue placing new bonds until the corresponding branch has been fully travelled by the double-folded path.
Once we return to the branching point, we continue along the direction which was not selected in the first instance.
Finally, we return to the origin and close the ring.
It is easy to see that this procedure ensures that each tree segment is visited exactly twice by oppositely oriented ring segments and that $\ell_{ring}=2N$ for the ring contour length.
For a given tree, there are $2^{n_3}$ possible wrappings, where $n_3$ is the total number of branching nodes of the tree.
Here we always consider averages over tree {\it ensembles}, where we analyse a single randomly generated wrapping per stored tree conformation.

%%%
\section{Results and discussion}\label{sec:Results}
%%%

\begin{figure}
\includegraphics[width=\columnwidth]{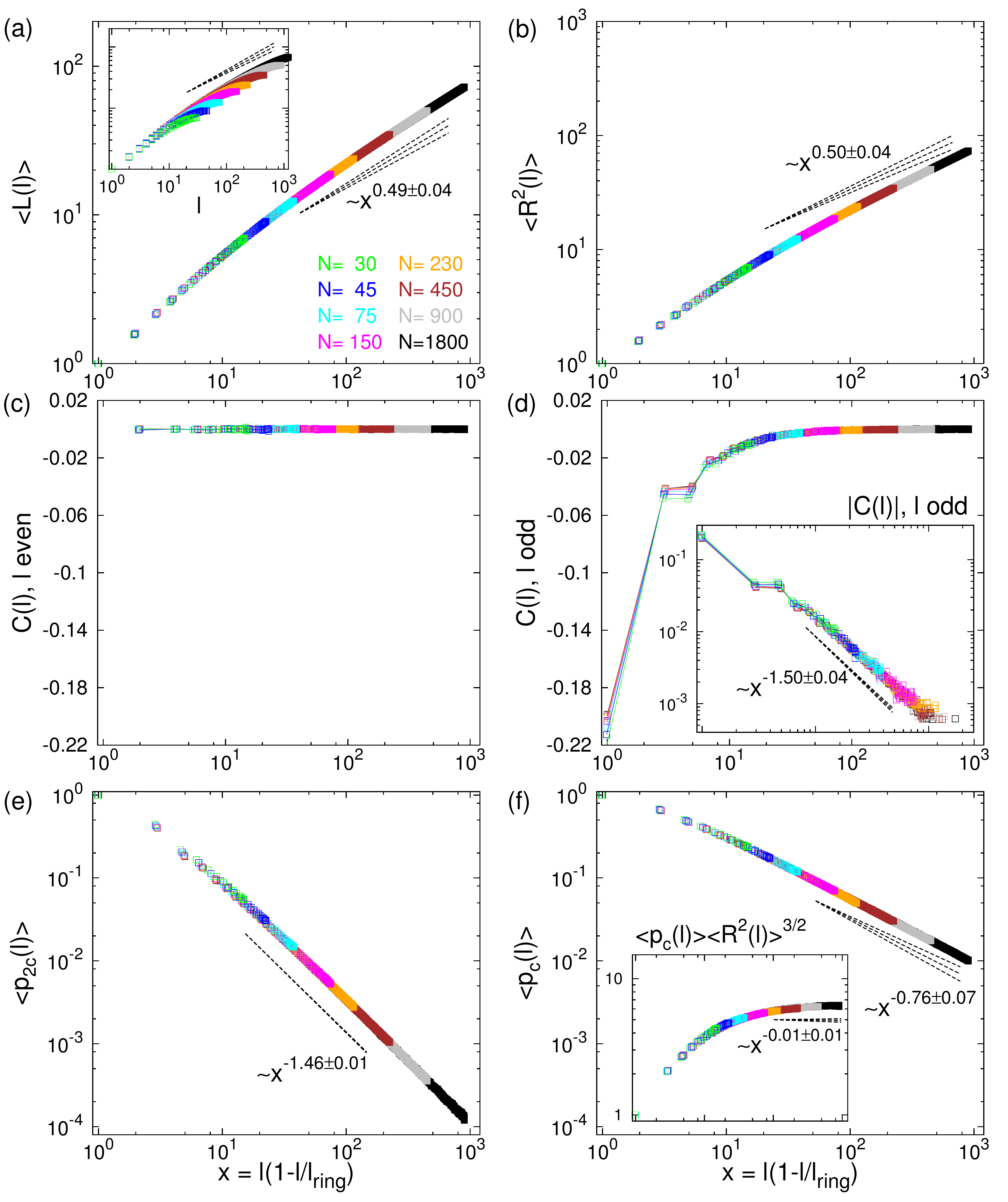}
\caption{
\label{fig:R2-vs-lring-IdTrees3d}
Conformational statistics of double-folded rings on $3d$ ideal randomly branching trees made of $N$ Kuhn segments.
(a)
$\langle L(\ell) \rangle \sim \ell^{\rho}$, average tree contour distance for ring contour distance $\ell$;
(b)
$\langle R^2(\ell) \rangle \sim \ell^{2\nu}$, end-to-end mean-square spatial distances;
(c, d)
$C(\ell) \sim \ell^{-2(1-\nu)}$, bond orientation correlation function.
In the insets here and in panels (c, d) in Fig.~\ref{fig:R2-vs-lring-SAT3d} to Fig.~\ref{fig:R2-vs-lring-Melt3d} data with relative errors $>20\%$ are considered below noise level and discarded from the plots;
(e)
$\langle p_{2c}(\ell) \rangle \sim \ell^{-(2-\epsilon)}$, average probability of secondary structure contacts.
The straight lines shown in these panels are calculated based on the average values and error bars of scaling exponents $\rho$, $\nu$ and $\epsilon$ presented in our works~\cite{RosaEveraers-JPA2016,RosaEveraers-JCP2016}.
(f)
$\langle p_{c}(\ell) \rangle \sim \ell^{-\nu(d+\theta_r)}$, contact probability.
The straight line is obtained by using the average value and error bars for the scaling exponent $\theta_r$ presented in this work, see Table~\ref{tab:ExpsSummaryTable}.
}
\end{figure}

\begin{figure}
\begin{center}
\includegraphics[width=\columnwidth]{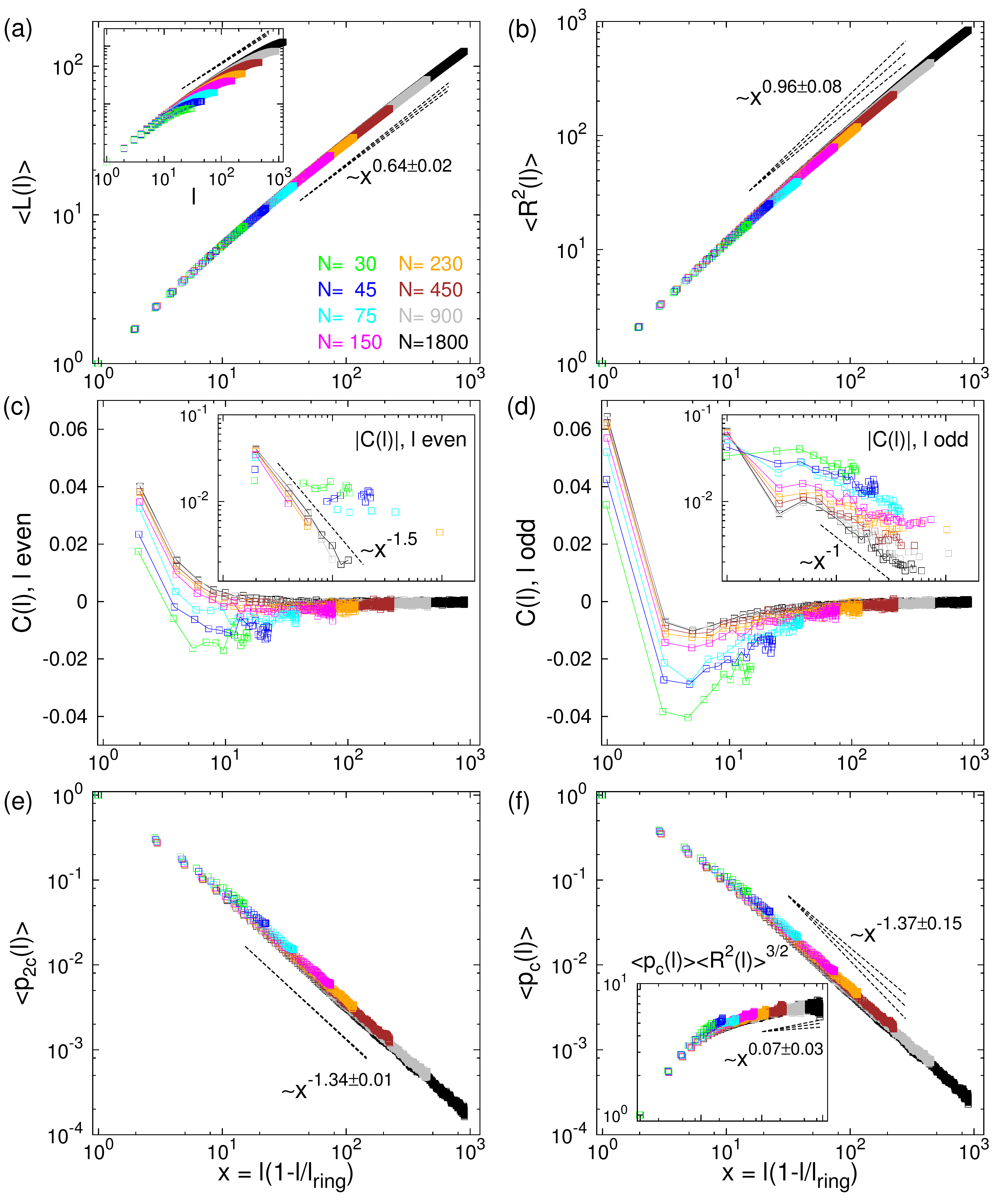}
\end{center}
\caption{
\label{fig:R2-vs-lring-SAT3d}
Conformational statistics of double-folded rings on $3d$ self-avoiding randomly branching trees made of $N$ Kuhn segments.
Notation and symbols are as in Fig.~\ref{fig:R2-vs-lring-IdTrees3d},
except for the {\it exact} straight lines $\sim \ell^{-1.5}$ and $\sim \ell^{-1}$ in the insets of panels (c) and (d).
}
\end{figure}

\begin{figure}
\begin{center}
\includegraphics[width=\columnwidth]{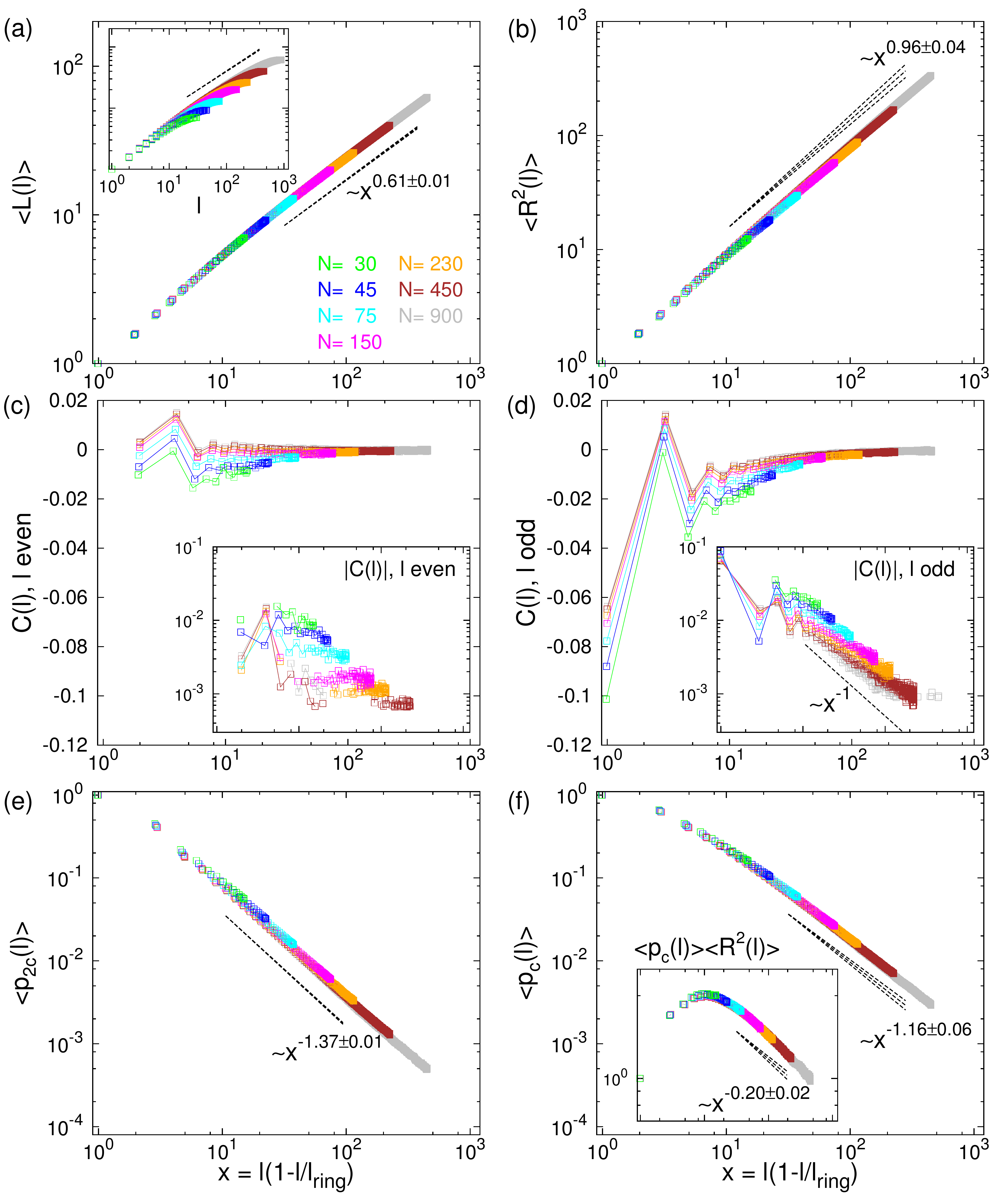}
\end{center}
\caption{
\label{fig:R2-vs-lring-Melt2d}
Conformational statistics of double-folded rings on $2d$ melt of randomly branching trees made of $N$ Kuhn segments.
Notation and symbols are as in Fig.~\ref{fig:R2-vs-lring-IdTrees3d},
except for the {\it exact} straight line $\sim \ell^{-1}$ in the inset of panel (d).
}
\end{figure}

\begin{figure}
\begin{center}
\includegraphics[width=\columnwidth]{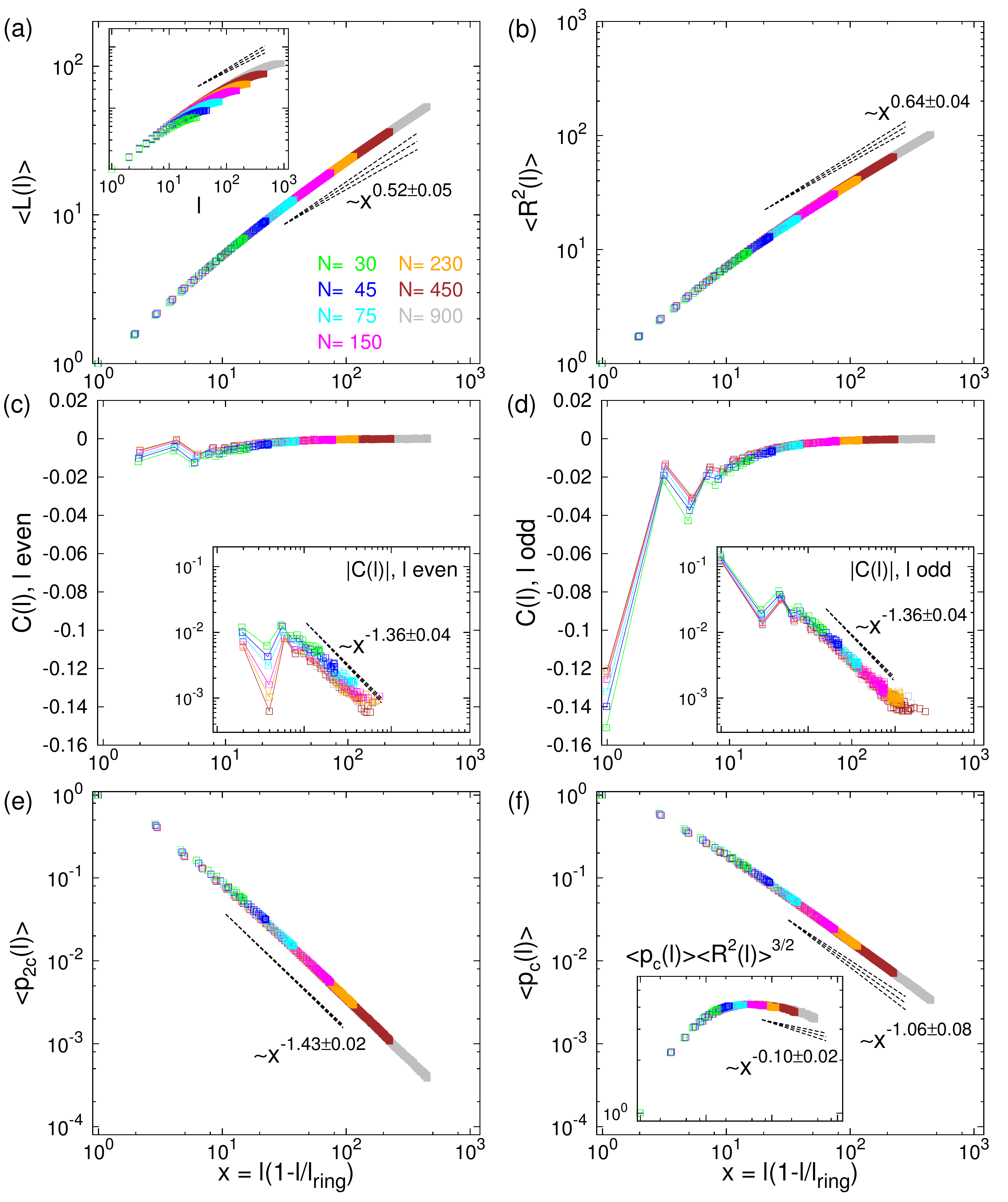}
\end{center}
\caption{
\label{fig:R2-vs-lring-Melt3d}
Conformational statistics of double-folded rings on $3d$ melt of randomly branching trees made of $N$ Kuhn segments.
Notation and symbols are as in Fig.~\ref{fig:R2-vs-lring-IdTrees3d}.
}
\end{figure}

%%%
\subsection{Tree contour distance as a function of ring contour distance}
%%%

The central quantity for understanding the conformational statistics of wrapped rings is the average tree contour distance, $L$, between two monomers as a function of their ring contour distance, $\ell$.
In general, $L\le\ell$. The minimal extension, $L=0$, corresponds to the case, where the ring wraps a branch of the tree composed of $\ell/2$ tree segments.
To reach the maximal extension, $L=\ell$, the ring has to follow a linear path of length $\ell$ on the tree. 
In general, the part of the tree wrapped by the ring section is in itself a tree composed of ${\cal O}(\ell)$ segments.
Adapting Eq.~(\ref{eq:rho}), we thus expect
\begin{equation}\label{eq:rho_ring}
\lim_{N\rightarrow\infty} \langle L(\ell) \rangle_N \sim \ell^\rho \, .
\end{equation}
Furthermore, due to the ring closure $\langle L(\ell) \rangle \equiv \langle L(\ell_{ring}-\ell) \rangle$ reaches its maximum for $\ell=\ell_{ring}/2$ before reducing to zero at the total ring size,
$\langle L(\ell_{ring}) \rangle\equiv0$.
The simplest functional form accounting for this constraint is
\begin{equation}\label{eq:rho_ring-constraint}
\langle L(\ell) \rangle_N \sim \left( \ell \left( 1 - \frac{\ell}{\ell_{ring}} \right) \right)^\rho \, .
\end{equation}
Panels (a) in Figs.~\ref{fig:R2-vs-lring-IdTrees3d} to \ref{fig:R2-vs-lring-Melt3d} show good agreement with this ansatz. Data are shown for ring contour distances up to $\ell_{ring}/2$; plotting them as a function of $\ell \left( 1 - \frac{\ell}{\ell_{ring}} \right)$ effectively reduces finite ring size effects (in the insets the same data are shown as a function of $\ell$). 
The dashed lines indicate the expected power law, where we used the effective exponents we extracted~\cite{RosaEveraers-JPA2016,RosaEveraers-JCP2016} for the trees underlying the present ring constructions.

%%%
\subsection{Secondary structure contacts}\label{sec:2aryContacts}
%%%

We define as a secondary structure contact a pair of monomers, which are neighbors on the tree, $L_{ij}\le l_K$, but not along the ring, $|j-i| > 1$ (Fig.~\ref{fig:DoubleFolding}(b)). We chose a finite contact distance to preserve the analogy with tertiary structure contacts (see below) and evaluated the contact probability $\langle p_{2c}(\ell) \rangle$ only for odd values of $\ell$ to avoid even/odd fluctuations induced by the lattice.
Modulo the finite contact radius, the probability to create a secondary structure contact at a ring distance $\ell$ equals (symbols {\it vs.} dashed lines) the probability to cut a branch of size $\ell/2$ from the tree:
$\langle p_{2c}(\ell) \rangle = p_{br}(\ell/2)$.
In particular, Eq.~(\ref{eq:branching probability}) suggests
\begin{eqnarray}
\label{eq:Contacts2aryStr}
\lim_{N\rightarrow\infty} \langle p_{2c}(\ell) \rangle_N &\sim& \ell^{-(2-\epsilon)} \, \\
\label{eq:Contacts2aryStr-constraint}
\langle p_{2c}(\ell) \rangle_N &\sim& \left( \ell \left( 1 - \frac{\ell}{\ell_{ring}} \right) \right)^{-(2-\epsilon)} \, .
\end{eqnarray}
Panels (e) in Figs.~\ref{fig:R2-vs-lring-IdTrees3d} to \ref{fig:R2-vs-lring-Melt3d} show that this is well supported by our data.

%%%
\subsection{Tree contour distance distributions}\label{sec:TreeContourPDFs}
%%%

The mean tree contour distance and the secondary structure contact probability can both be obtained from the full tree contour distance distribution, $p_N(L|\ell)$, for ring sections of length $\ell$ on rings composed of $2N$ Kuhn segments:
\begin{eqnarray}
\langle L(\ell) \rangle_N &=& \int_0^L L\, p_N(L|\ell) \, dL\\
\langle p_{2c}(\ell) \rangle_N &=& \int_0^{l_K} p_N(L|\ell) \, dL \label{eq:p2c of pL}
\end{eqnarray}

\begin{figure}
%\begin{center}
\includegraphics[width=0.5\textwidth]{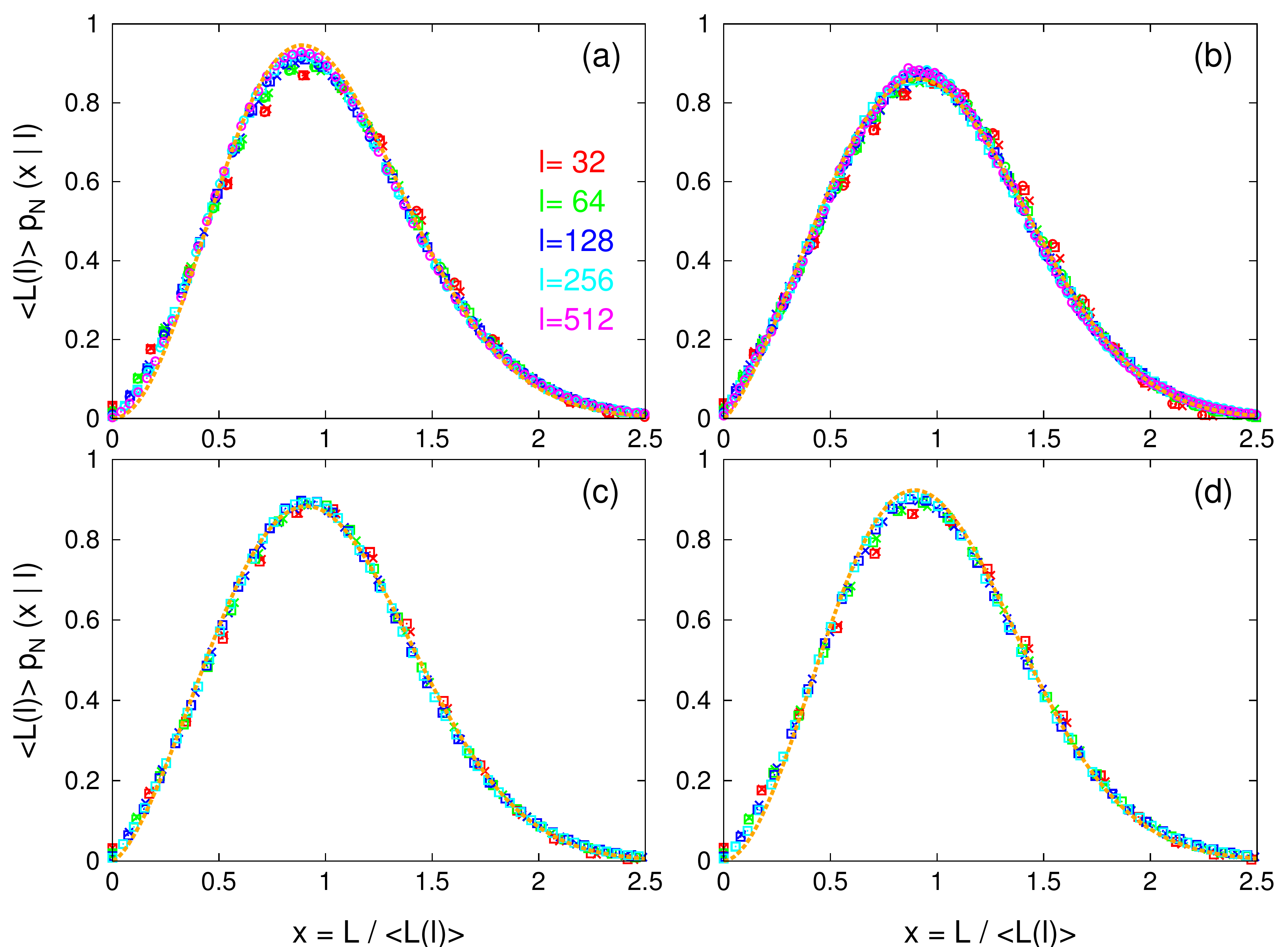}
%\end{center}
\caption{
\label{fig:LtreesPDFs}
Rescaled probability distributions, $p_{N}(L | \ell)$, Eq.~(\ref{eq:RdC-LTree-Lring}), of tree contour distances, $L$, for given ring contour distance $\ell$.
Different symbols correspond to different tree sizes: ($\times$) $N=450$, ($\square$) $N=900$, ($\circ$) $N=1800$.
Results for:
(a) $3d$ ideal trees;
(b) $3d$ self-avoiding trees;
(c) $2d$ melt of trees;
(d) $3d$ melt of trees.
Orange lines in panels (a-d) are for best fits (see Table~\ref{tab:ExpsSummaryTable}) to the Redner-des Cloizeaux (RdC) function, Eq.~(\ref{eq:q_RdC_L}). 
}
\end{figure}

Figure~\ref{fig:LtreesPDFs} shows, that the measured contour distance distributions
fall onto universal master curves, when plotted as a function of the rescaled contour distance
$x = L / \langle L(\ell) \rangle_N$:
\begin{equation}\label{eq:RdC-LTree-Lring}
p_N(L|\ell) = \frac1{\langle L(\ell) \rangle_N}\  q \left(\frac{L}{\langle L(\ell) \rangle_N}\right) \, .
\end{equation}
These master curves are well described by the one-dimensional Redner-des Cloizeaux (RdC) form (orange lines in Fig.~\ref{fig:LtreesPDFs}):
\begin{equation}\label{eq:q_RdC_L}
q(x) = C_L \, x^{\theta_L}\  \exp \left( -(K_L x)^{t_L} \right) \, .
\end{equation}
The constants
\begin{eqnarray}
C_L &=& t_L \, \frac{\Gamma^{\theta_L+1}((\theta_L+2)/t_L)}{\Gamma^{\theta_L+2}((\theta_L+1)/t_L)}
\label{eq:RdC_C_L}\\
K_L &=& \frac{\Gamma((\theta_L+2)/t_L)}{\Gamma((\theta_L+1)/t_L)}
\label{eq:RdC_K_L}
\end{eqnarray}
follow from the conditions that $p_N(L|\ell)$ is normalized to $1$ and that the first moment, $\langle L(\ell) \rangle_N$, is the only relevant scaling variable.
Estimated values for $(\theta_L, t_L)$ in the asymptotic ($N\rightarrow\infty$) limit of large trees are summarized in Table~\ref{tab:ExpsSummaryTable}.
More details concerning best fits of Eq.~(\ref{eq:q_RdC_L}) to data for specific values of $N$ and large-$N$ extrapolations of scaling exponents are given in the Appendices and in Table~\ref{tab:ThetaLTL-Fits}.

\begin{table*}
\begin{tabular}{cccccccccc}
\hline
\hline
\\
& Relation to & \multicolumn{2}{c}{$3d$ ideal trees} & \multicolumn{2}{c}{$3d$ self-avoiding trees} & \multicolumn{2}{c}{$2d$ melt of trees} & \multicolumn{2}{c}{$3d$ melt of trees} \\
& other exponents & (a) & (b) & (a) & (b) & (a) & (b) & (a) & (b) \\ 
\hline
\\
$\theta_L$ & $\frac{2}{\rho}-2$ & $2.05 \pm 0.09$ & $2.08\pm0.34$ & $1.33 \pm 0.06$ & $1.13\pm0.10$ & $1.47 \pm 0.06$ & $1.26\pm0.04$ & $1.82 \pm 0.06$ & $1.85\pm0.37$ \\
\\
$t_L$ & $\frac{1}{1-\rho}$ & $2.01 \pm 0.11$ & $1.96\pm0.15$ & $2.51 \pm 0.09$ & $2.78\pm0.15$ & $2.43 \pm 0.08$ & $2.58\pm0.05$ & $2.15 \pm 0.08$ & $2.08\pm0.22$ \\
\\
$\theta_r$ & $\min( \theta_{path} , \frac{2-\rho}{\nu} - d )$ & $0.02\pm0.05$ & $0$ & $-0.14\pm0.07$ & $-0.17\pm0.28$ & $0.42\pm0.02$ & $0.63\pm0.04$ & $0.31\pm0.05$ & $0.28\pm0.01$ \\
\\
$t_r$ & $\frac{1}{1-\nu}$ & $1.34\pm0.07$ & $1.33\pm0.04$ & $2.00\pm0.06$ & $1.92\pm0.15$ & $1.90\pm0.06$ & $1.92\pm0.07$ & $1.53\pm0.02$ & $1.47\pm0.04$ \\
\\
\hline
\hline
\end{tabular}
\caption{
\label{tab:ExpsSummaryTable}
Scaling exponents $(\theta_L, t_L)$ and $(\theta_r, t_r)$ for RdC distribution functions, Eqs.~(\ref{eq:q_RdC_L}) and~(\ref{eq:q_RdC_r}).
Columns denoted by (a) and (b) correspond to, respectively,
asymptotic values in the infinite ($N\rightarrow\infty$) tree limit
and
after substitution of values for scaling exponents $\rho$, $\nu$ and $\theta_{path}$ obtained in Refs.~\cite{RosaEveraers-JPA2016,RosaEveraers-JCP2016,RosaEveraers-PRE2017} into the relations summarized in this table.
For details on the derivation, see Tables~\ref{tab:ThetaLTL-Fits} and~\ref{tab:ThetaRingTRing-Fits} in the Appendices.
}
\end{table*}

As for the corresponding path length distribution for trees~\cite{RosaEveraers-PRE2017}, we can give a physical interpretation of the observed (effective) exponents.
For small tree contour distances, $x\rightarrow0$, Eq.~(\ref{eq:q_RdC_L}) reduces to $q(x) = C_L \, x^{\theta_L}$. Inserting into Eq.~(\ref{eq:p2c of pL}) for the secondary structure contact probability and equating with Eq.~(\ref{eq:Contacts2aryStr}) yields:
\begin{equation}\label{eq:ThetaLvsRho}
\theta_L = \frac2\rho-2 \, .
\end{equation}
Using the numerical estimates for $\rho$ from Refs.~\cite{RosaEveraers-JPA2016,RosaEveraers-JCP2016},
Eq.~(\ref{eq:ThetaLvsRho}) is in agreement with the extrapolated values for $\theta_L$ in the limit $\ell\rightarrow\infty$, see Table~\ref{tab:ExpsSummaryTable} and the Appendices for numerical details.
Moreover, note that $\theta_L$ differs from the corresponding exponent $\theta_l=\frac1\rho-1$ for the contour distance distribution on trees introduced in Ref.~\cite{RosaEveraers-PRE2017}.

To estimate the probability for observing large contour distances, $L$, we can follow Ref.~\cite{RosaEveraers-PRE2017} and formulate the problem in terms of Pincus blobs~\cite{PincusBlob1976}.
Ring sections, that follow an almost linear path on the tree, behave as if they would wrap a string of $\ell/g$ unperturbed trees of size $\xi\sim g^\rho$. This suggests that
\begin{equation}\label{eq:TLvsRho}
t_L = \frac{1}{1-\rho}\ .
\end{equation}
As for the exponents $\theta_L$, this equation is in reasonable agreement with numerical extrapolations of $t_L$ in the limit $\ell\rightarrow\infty$, see Table~\ref{tab:ExpsSummaryTable} and the Appendices for numerical details.

%%%
\subsection{Mean square spatial distances along the ring}
%%%

Having obtained a complete characterization of wrapping on the level of the connectivity graph characterising the tree, we can now turn our attention to the spatial embedding of the wrapped ring. 
The simplest measure to consider is the mean-square spatial distance, $\langle R^2(\ell) \rangle_N$, of monomers separated by a contour distance $\ell$ along the ring.
Combining Eqs.~(\ref{eq:nu_path}) and~(\ref{eq:rho_ring-constraint}) with the relation $\nu=\rho \, \nu_{\rm path}$ suggests
\begin{equation}\label{eq:NuDef-constraint}
\langle R^2(\ell) \rangle_N \sim \left( \ell \left( 1 - \frac{\ell}{\ell_{ring}} \right) \right)^{2\nu} \, .
\end{equation}
This equation is supported by corresponding numerical results (symbols {\it vs.} dashed lines) shown in panels (b) in Figs.~\ref{fig:R2-vs-lring-IdTrees3d} to \ref{fig:R2-vs-lring-Melt3d}, where
$\nu=1/4$ for ideal trees,
$\nu=1/2$ for self-avoiding trees in $d=3$ dimensions, and
$\nu=1/d$ for tree melts in $d$ dimensions.

%%%
\subsection{Bond orientation correlation function}
%%%

Additional insight into the polymer conformation can be gained by considering the bond orientation correlation function, $C(\ell) = \langle \vec b_i \cdot \vec b_{i+\ell} \rangle / |\vec b|^2$.
$C(\ell)$ is related to the mean-square distances via the identity:
\begin{eqnarray}
\langle R^2(\ell) \rangle 
&=& 2 \int_0^\ell \int_s^\ell C(s'-s)\, ds'ds 
\nonumber\\
&=& 2 \int_0^\ell (\ell-\delta) C(\delta)\, d\delta
\label{eq:Rsqr_cont}
\end{eqnarray}
and
\begin{equation}\label{eq:BACF of R2}
C(\ell) =  \frac12 \frac {d^2}{d \ell^2}\langle R^2(\ell) \rangle \ .
\end{equation}
For $\nu\ne1/2$, Eq.~(\ref{eq:BACF of R2}) implies
\begin{equation}\label{eq:nu of BACF}
C(\ell) \sim 2\nu(2\nu-1)\ell^{2(\nu-1)}\ .
\end{equation} 
In particular, $C(\ell)<0$ for $\nu<1/2$.
The case $\nu=1/2$ is special.
The best-know examples are:
$C(\ell)\equiv0$ for $\ell>0$ (Kuhn model or freely jointed chain~\cite{RubinsteinColby}) and
$C(\ell)\equiv \exp(-2\ell/l_K)$ (worm-like chain~\cite{KratkyPorod}).
However, Eqs.~\ref{eq:Rsqr_cont} and~\ref{eq:BACF of R2} are also compatible with a power-law decay, $C(\ell)\sim \ell^{-\omega}$, as long as $\omega>1=(2-2\nu)$.
An interesting example are long-range bond orientation correlations in polymer melts~\cite{WittmerPRL2004}, where $\omega=3/2$.

For tightly wrapped rings of the type considered here, it turns out to be useful to distinguish even and odd ring contour distances.
Consider first ideal trees, where there is no orientation correlation between different {\em tree} segments.
As ring segments with arbitrary even $\ell=|j-i|>0$ can never be co-localized on the same tree segment, $C(\ell) \equiv 0$ for even $\ell$ (panel (c) in Fig.~\ref{fig:R2-vs-lring-IdTrees3d}).
In contrast, ring segments with odd $\ell=|j-i|$ located on the {\em same} tree segment are {\em anti-}correlated.
Using Eq.~(\ref{eq:Contacts2aryStr}) this implies $C(\ell) \sim -\ell^{-(2-\epsilon)} = -\ell^{-3/2}$ for odd $\ell$ (panel (d) in Fig.~\ref{fig:R2-vs-lring-IdTrees3d}).
This power law is compatible with Eq.~(\ref{eq:nu of BACF}) as $\nu=1/4$ for ideal trees~\cite{ZimmStockmayer49}.

For interacting trees, the bond orientation correlation function exhibits corresponding even/odd fluctuations (panels (c) and (d) in Figs.~\ref{fig:R2-vs-lring-SAT3d} to \ref{fig:R2-vs-lring-Melt3d}).
In qualitative agreement with the ideal case, distant bonds along the ring have a tendency to be anti-correlated, $C(\ell)<0$ for $\ell \gg 1$.
For rings wrapped around trees from $3d$ melts with $\nu=1/3$, our results are again in agreement with the power law decay expected from Eq.~(\ref{eq:nu of BACF}):
$C(\ell) \sim -\ell^{2(\nu-1)} = - \ell^{-4/3}$.
The bond orientation correlation functions for rings wrapped around trees from $2d$ melts as well as self-avoiding trees with $\nu=1/2$ 
are compatible with a convergence to an asymptotic power law decay, $C(\ell) \sim \ell^{-\omega}$, with an exponent, $\omega \approx 1$, close to the limiting value.

%%%
\subsection{Tertiary structure contacts and total contacts}\label{sec:3aryContacts}
%%%

We define as a tertiary structure contact a pair of monomers, which are neighbors in space, $|\vec r_{ij}| \le l_K$, but neither on the tree, $L_{ij}> l_K$, nor along the ring, $|j-i| > 1$ (Fig.~\ref{fig:DoubleFolding}(b)). 
Clearly, the sum of both, secondary (see Sec.~\ref{sec:2aryContacts}) and tertiary contact probabilities, produces the total contact probability, $\langle p_c(\ell) \rangle$, between pair of ring monomers at contour distance $\ell$.
Our data for $\langle p_c(\ell) \rangle$ for the various tree types are shown in panels (f) in Figs.~\ref{fig:R2-vs-lring-IdTrees3d} to \ref{fig:R2-vs-lring-Melt3d}.
As for secondary structure contacts probabilities, Eq.~(\ref{eq:Contacts2aryStr-constraint}),
we can take into account the ring closure constraint through the ansatz
\begin{equation}\label{eq:GammaDef-constraint}
\langle p_{c}(\ell) \rangle_N  \sim  \left( \ell \left( 1-\frac{\ell}{\ell_{ring}} \right) \right)^{-\gamma_r} \, ,
\end{equation}
which effectively reduces finite ring-size effects.

%%%
\subsection{End-to-end distance distributions for ring sections}\label{sec:End-to-End-PDFs}
%%%

The mean square internal distances and the contact probability can both be obtained from the full end-to-end distance distribution, $p_N(\vec r|\ell)$, for ring sections of length $\ell$ on rings composed of $2N$ Kuhn segments:
\begin{eqnarray}
\langle R^2(\ell) \rangle_N &=& \int_0^\infty |\vec r|^2 \, p_N(\vec r|\ell) \, d\vec r\\
\langle p_{c}(\ell) \rangle_N &=& \int_0^{l_K} p_N(\vec r|\ell) \, d\vec r \label{eq:pc of pr}
\end{eqnarray}

\begin{figure}
%\begin{center}
\includegraphics[width=0.5\textwidth]{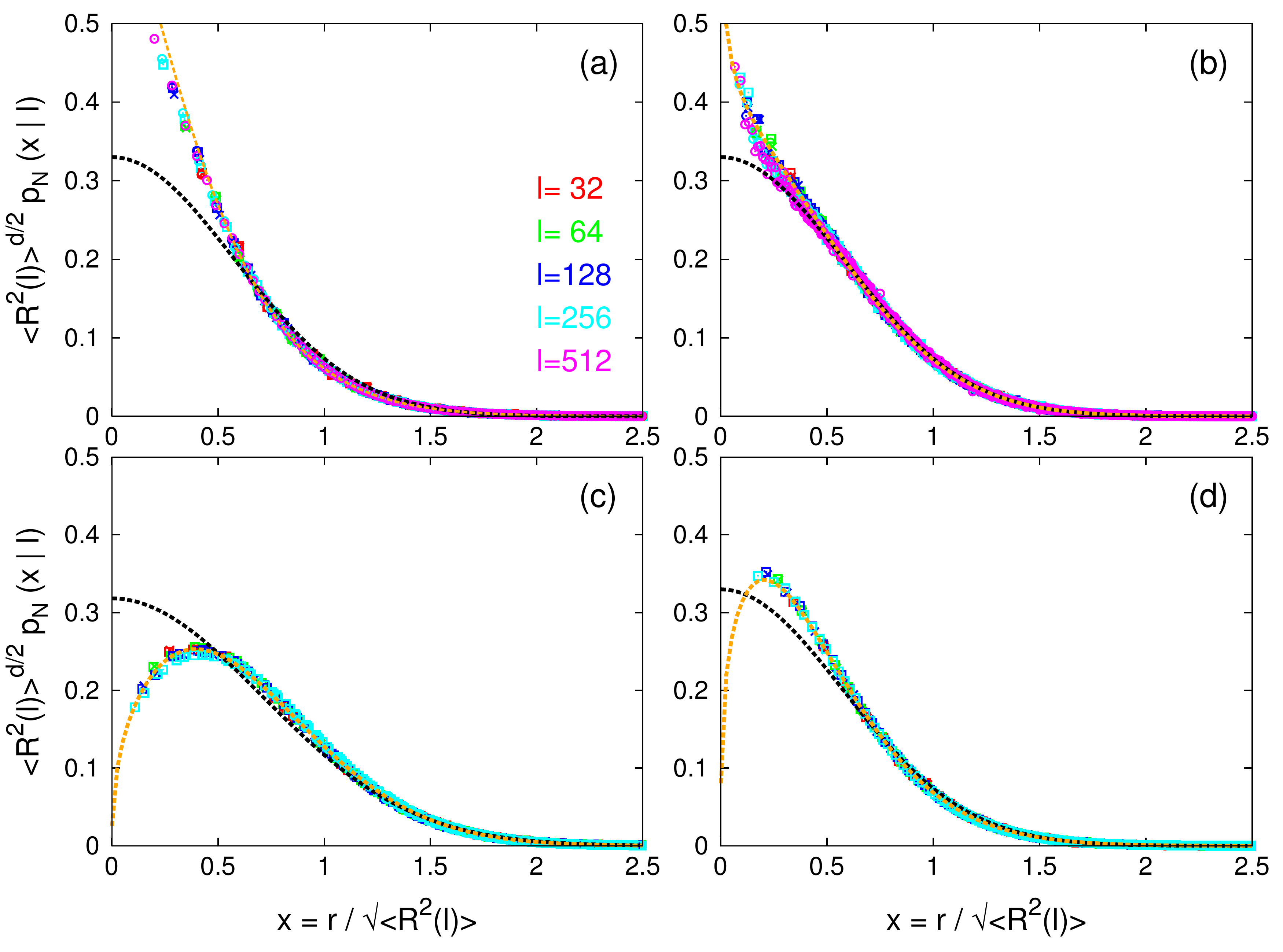}
%\end{center}
\caption{
\label{fig:IntDistsPDFs}
Probability distribution functions, $p_{N}({\vec r} | \ell)$, of end-to-end spatial distances, $r$, for given ring contour distance $\ell$.
Notation and symbols are as in Fig.~\ref{fig:LtreesPDFs}.
Orange lines in panels (a-d) are for best fits (see Table~\ref{tab:ExpsSummaryTable}) to the RdC function, Eq.~(\ref{eq:q_RdC_r}).
Black lines are for the $d$-dimensional Gaussian function.
}
\end{figure}

Figure~\ref{fig:IntDistsPDFs} shows, that the measured end-to-end distance distributions
fall onto universal master curves, when plotted as a function of the rescaled distance $\vec x =  \vec r / \sqrt{\langle R^2(\ell) \rangle_N}$:
\begin{equation}\label{eq:RdC-Lring-E2E}
p_N( \vec r | \, \ell) = \frac1{\langle R^2(\ell) \rangle_N^{d/2}}\  q\left(\frac{\vec r}{ \sqrt{\langle R^2(\ell) \rangle_N}}\right) \, .
\end{equation}
These master curves are well described by the $d$-dimensional Redner-des Cloizeaux (RdC) form (orange lines in Fig.~\ref{fig:IntDistsPDFs}):
\begin{equation}\label{eq:q_RdC_r}
q(\vec x) = C \, |\vec x|^{\theta_{r}}\  \exp \left( -(K |\vec x|)^{t_{r}} \right)
\end{equation}
with the constants $C$ and $K$ given by Eqs.~(\ref{eq:RdC_C}) and (\ref{eq:RdC_K}).
In particular, the contact exponent defined in Eq.~(\ref{eq:GammaDef-constraint}) is given by
\begin{equation}\label{eq:Gamma vs theta_r}
\gamma_r=\nu(d+\theta_r) \, .
\end{equation}
Estimated values for $(\theta_r, t_r)$ in the asymptotic ($N\rightarrow\infty$) limit of large trees are given in Table~\ref{tab:ExpsSummaryTable}.
More details concerning best fits of Eq.~(\ref{eq:q_RdC_r}) to data for specific values of $N$ and large-$N$ extrapolations of scaling exponents are given in the Appendices and in Table~\ref{tab:ThetaRingTRing-Fits}.

In the following, we relate the characteristic exponents $\theta_{r}$ and $t_{r}$ to our previous results by using Eqs.~(\ref{eq:q_RdC_path}) and (\ref{eq:q_RdC_L}) together with the identity 
\begin{equation}\label{eq:pr_convolution}
p_N(\vec r | \ell) = \int_0^{\infty} p_N(\vec r|L) \, p_N(L | \ell) \, d L\ \ ,
\end{equation}
which states that the probability for reaching a particular spatial distance $\vec r$ at given ring contour distance $\ell$ can be calculated by adding up the contributions from all possible tree contour distances, $0\le L\le \ell$. 
The behavior of $p_N(\vec r | \ell)$ for large distances, $r >\sqrt{\langle R^2(\ell) \rangle_N}$, can be estimated from the contour distance $L^\ast(r)$, which makes the dominant contribution to particle pairs found at the spatial distance $r$.
Combining the arguments of the compressed exponentials in Eqs.~(\ref{eq:q_RdC_path}) and~(\ref{eq:q_RdC_L}), this requires the minimization of
$\left( \frac{L}{\langle L(\ell) \rangle_N} \right)^{t_L} + \left( \frac{r}{\sqrt{\langle R^2(L) \rangle_N}} \right)^{t_{path}}$
and yields
\begin{equation}\label{eq:tTree}
t_{r} = \frac{t_L \, t_{path}} {t_L + t_{path} \, \nu_{path}} = \frac{1}{1-\nu} \, .
\end{equation}
By using the numerical estimates for $\nu$ from our works~\cite{RosaEveraers-JPA2016,RosaEveraers-JCP2016}, Eq.~(\ref{eq:tTree}) is in good agreement with the extrapolated values for $t_r$ (see Table~\ref{tab:ExpsSummaryTable} and the Appendices for numerical details).

In the opposite limit of small distances, $r<\sqrt{\langle R^2(\ell) \rangle_N}$, the exponentials in Eqs.~(\ref{eq:q_RdC_path}) and (\ref{eq:q_RdC_L}) can be set equal to one in between $L_{min}\sim r^{1/\nu_{path}}$ and $L_{max}\sim L^\rho$.
Contributions to the integral Eq.~(\ref{eq:pr_convolution}) from outside of the interval $[L_{min},L_{max}]$ are negligible,
so that the integral is of the form $\int_{L_{min}}^{L_{max}} L^{-\alpha} dL \stackrel{\alpha\ne1}{=} (\alpha-1) \left(L_{min}^{1-\alpha} -  L_{max}^{1-\alpha}  \right)$ with 
$\alpha=\nu_{path} (\theta_{path}+d)-\theta_L$. Depending on the value of $\alpha$, the integral is dominated by the lower or the upper cutoff for $L$.

For $\alpha<1$, the integral is dominated by contributions from long paths with $\langle R^2(L) \rangle_N\gg r^2$.
%{\bf
By using values for $\nu_{path}$ and $\theta_{path}$ from our works~\cite{RosaEveraers-JPA2016,RosaEveraers-JCP2016,RosaEveraers-PRE2017} and $\theta_L$ from this work (Table~\ref{tab:ExpsSummaryTable}),
this is the case for 
rings wrapping ideal trees ($\alpha = -0.52\pm0.12$),
rings wrapping trees from $2d$ and $3d$ melts ($\alpha = 0.58\pm0.10$ and $\alpha = 0.13\pm0.09$, respectively).
%}
The only $r$-dependence comes through the explicit $r^{\theta_{path}}$ term and hence
\begin{equation}
\theta_{r} =  \theta_{path} \ \ \ \ \mbox{if\ \ \ } \theta_{path}< \frac{2-\rho}\nu-d\ .
\end{equation}
This is borne out by our data (symbols {\it vs.} dashed lines in panels (f) in Figs.~\ref{fig:R2-vs-lring-IdTrees3d}, \ref{fig:R2-vs-lring-Melt2d}, and \ref{fig:R2-vs-lring-Melt3d}).
In these cases, the total contact probability exceeds by far the secondary structure contact probability ({\it cf.} the corresponding panels (e)).

\begin{figure}
\begin{center}
\includegraphics[width=0.5\textwidth]{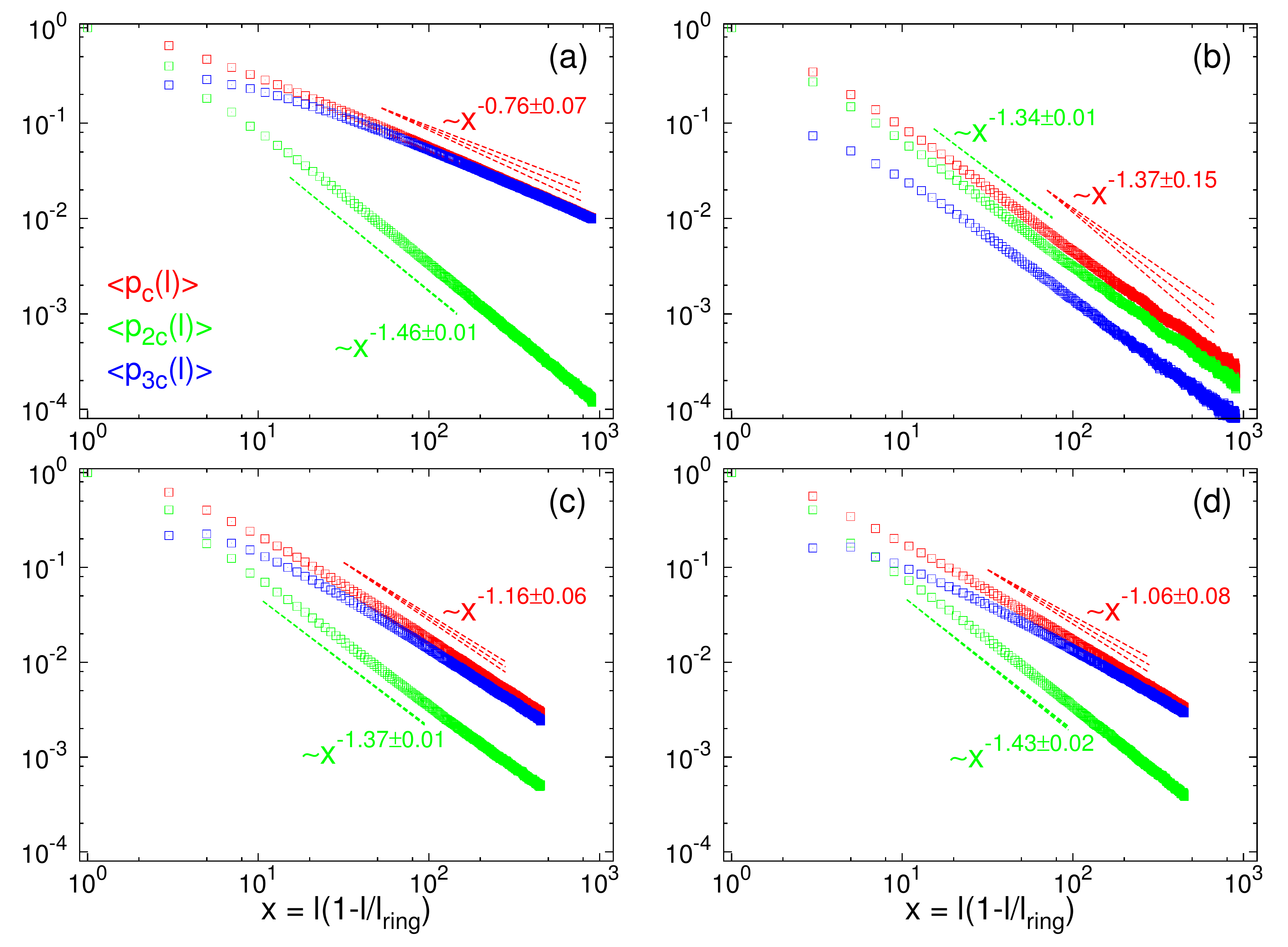}
\end{center}
\caption{
\label{fig:SecondaryTertiaryContacts}
Secondary structure ($\langle p_{2c}(\ell) \rangle$),
tertiary structure ($\langle p_{3c}(\ell) \rangle$)
and
generic contact probabilities ($\langle p_{c}(\ell) \rangle = \langle p_{2c}(\ell) \rangle + \langle p_{3c}(\ell) \rangle$)
between ring monomers for contour length separation $\ell$ (see Fig.~\ref{fig:DoubleFolding}(b) for definitions).
Dashed lines are for the expected scaling behaviors
$\langle p_{2c}(\ell) \rangle \sim \ell^{-(2-\epsilon)}$
and
$\langle p_{c}(\ell) \rangle \sim \ell^{-\nu(d+\theta_r)}$,
as in panels (e) and (f) of Figs.~\ref{fig:R2-vs-lring-IdTrees3d} to~\ref{fig:R2-vs-lring-Melt3d}.
Results for:
(a) $3d$ ideal trees ($N=1800$);
(b) $3d$ self-avoiding trees ($N=1800$);
(c) $2d$ melt of trees ($N=900$);
(d) $3d$ melt of trees ($N=900$).
}
\end{figure}

In the opposite limit, short paths dominate and the total contact probability is essentially given by the secondary structure contact probability.
Of the systems we have studied, only rings wrapped around self-avoiding trees in $d=3$ dimensions fall into this category ($\alpha=1.68\pm0.20$, panels (e) and (f) in Figs.~\ref{fig:R2-vs-lring-SAT3d}). In this case,
\begin{equation}
\theta_{r} =  \frac{2-\rho}{\nu} - d \ \ \ \ \mbox{if\ \ \ }  \frac{2-\rho}\nu-d<\theta_{path}\ 
\end{equation}
in agreement with Eqs.~(\ref{eq:Contacts2aryStr}) and~(\ref{eq:Gamma vs theta_r}).
Summarizing,
\begin{equation}\label{eq:ThetaTree}
\theta_{r} = \min( \theta_{path} , \frac{2-\rho}{\nu} - d ) \, .
\end{equation}
Once again Eq.~(\ref{eq:ThetaTree}) is in good agreement with the fitted values for $\theta_r$ (see Table~\ref{tab:ExpsSummaryTable} and the Appendices for numerical details), if we use the numerical values for $\rho$, $\nu$ and $\theta_{path}$ from Refs.~\cite{RosaEveraers-JPA2016,RosaEveraers-JCP2016,RosaEveraers-PRE2017}.
However, as illustrated by Fig.~\ref{fig:SecondaryTertiaryContacts}, 
the prefactors of these power laws also matter.
For rings wrapped around trees from, in particular, $3d$ melts (panel (d)), the relative weight of the contributions of secondary and tertiary structure contacts to $\langle p_{c}(\ell) \rangle_N$ is a function of ring contour distance, $\ell$, giving rise to a small crossover with contact probabilities being described by intermediate effective exponents.

%

%%%
\section{Conclusions}\label{sec:Concls}
%%%

The mapping to lattice trees~\cite{KhokhlovNechaev85,RubinsteinPRL1986} presents an elegant simplification for a polymer problem that is otherwise difficult to treat: the conformation of ring polymers constrained by the absence of topological links with each other or a lattice of obstacles representing the gels used in electrophoresis. The statistical physics of lattice trees can be studied within Flory theory~\cite{RosaEveraers-SoftMatter2017}. Computer simulations and scaling arguments allow to refine the values of exponents and to explore the distribution functions characterising tree behavior beyond the average regime~\cite{RosaEveraers-PRE2017}. 

The results we have presented in this article show, that it is relatively straightforward (albeit not completely trivial) to transfer results obtained for trees to the original ring systems. ``Navigating'' on the tree along a wrapped ring mixes the two basic concepts used for  characterizing trees. 
The branching statistics of the trees controls the central quantity for understanding the conformational statistics of wrapped rings: the increase of the tree contour distance, $\langle L(\ell) \rangle \sim \ell^\rho$, between two monomers as a function of their ring contour distance, $\ell$. 
Spatial distances depend on the conformations of linear paths on the tree in the embedding space.
Using $\langle R^2(L) \rangle \sim L^{2\nu_{path}}$ with $\nu=\rho \nu_{path}$ for trees, one recovers for wrapped rings the familiar relation  $\langle R^2(\ell) \rangle \sim \ell^{2\nu}$.
The ring closure constraint is effectively dealt with by expressing observables as a function of $\ell (1-\ell/\ell_{ring})$.
Interestingly, the distribution functions $p(\vec r | \ell)$ and $p(L | \ell)$ for the spatial distance, $\vec r$, and tree contour distance, $L$, between monomers show excellent scaling behavior when expressed as a function of reduced spatial and contour distance, $\vec r/\sqrt{\langle R^2(\ell) \rangle}$ and $L / \langle L(\ell) \rangle$.
As for trees~\cite{RosaEveraers-PRE2017}, the scaled distributions turn out to be of the Redner-des Cloizeaux form~\cite{Redner1980,DesCloizeauxBook}.
They are characterized by two exponents, $t$ and $\theta$, which control the small and the large-scale behavior and which can be related to the other exponents describing rings and the trees they are wrapped around.

%%%%%%%%%
%\bibliographystyle{unsrt}
%\bibliography{biblio}

%%%%%%%%%

%%%
\appendix
%%%

%%%
\section{Model and methods}\label{sec:ModelMethods}
%%%

%In Section~\ref{sec:TreeEnsembles} we describe briefly the computer algorithm used for generating equilibrated conformations of randomly branching trees.
%The interested reader may look for additional details into our former works~\cite{RosaEveraers-JPA2016,RosaEveraers-JCP2016,RosaEveraers-PRE2017}.
%In Sec.~\ref{sec:DoubleFoldedRingsConstruction} we describe the algorithm for constructing double-folded rings on trees and
%In Sec.~\ref{sec:ScalExpsRdC} we discuss how to extract scaling exponents for RdC functions and contacts.
%Quantitative details as well as tabulated values for single-tree statistics are presented in the Supplemental Material.

%%%
\subsection{Generation of equilibrated conformations of randomly branching trees}\label{sec:TreeEnsembles}
%%%

Spatial conformations of branched polymers were generated according to the lattice polymer model extensively described in our former works~\cite{RosaEveraers-JPA2016,RosaEveraers-JCP2016,RosaEveraers-PRE2017}.
Full details can be found there, here we just provide a concise description of the method.

Briefly, we simulated randomly branching polymers with volume interactions
by employing a slightly modified version of the ``amoeba'' Monte Carlo algorithm~\cite{SeitzKlein1981} for trees on the cubic lattice with periodic boundary conditions.
In this model, connected nodes occupy adjacent lattice sites.
As there is no bending energy term, the lattice constant equals the Kuhn length, $l_K$, of linear paths across ideal trees.
The functionality of nodes is restricted to the values $f=1$ (a leaf or branch tip), $f=2$ (linear chain section), and $f=3$ (branch point).
Here, we consider $3d$ single trees in good solvent (self-avoiding trees), and $2d$ and $3d$ trees in melt with Kuhn density $\rho_K l_K^d=2$.
Tree sizes are in the range $30 \leq N \leq 1800$ for ideal and self-avoiding trees and $30 \leq N \leq 900$ for trees in melt.

%%%
\subsection{Numerical fitting procedures for deriving scaling exponents for Redner-des Cloizeaux functions}\label{sec:ScalExpsRdC}
%%%

Single estimates of scaling exponents $(\theta_L, t_L)$ and $(\theta_r, t_r)$ for Redner-des Cloizeaux (RdC) functions Eq.~(\ref{eq:q_RdC_L}) and Eq.~(\ref{eq:q_RdC_r}) describing the behavior of distribution functions $p_N(L|\ell)$ and $p_N({\vec r}|\ell)$ (Figs.~\ref{fig:LtreesPDFs} and~\ref{fig:IntDistsPDFs}, lines {\it vs.} symbols respectively) are obtained through the following procedure.
First,
we fit the analytical expressions for the RdC functions to our measured distributions for the largest tree sizes ($N=450, 900, 1800$) and for specific values of ring contour distances $\ell = 32, 64, ..., \approx N/4$.
Corresponding results for $(\theta_L, t_L)$ and $(\theta_r, t_r)$ are summarized in Tables~\ref{tab:ThetaLTL-Fits} and~\ref{tab:ThetaRingTRing-Fits}, respectively.

As, in general, the effective exponents $\theta_L$ and $t_L$ show consistent finite-size effects depending monotonously on $\ell$, we have used the empirical expression:
\begin{equation}\label{eq:FiniteSizeCorrectsFit}
\theta_L(\ell) = \theta_L(\infty) + A \, \ell^{-\Delta}
\end{equation}
containing free parameters $\theta_L(\infty)$, $A$ and $\Delta$
and an analogous one for $t_L(\ell)$ to estimate asymptotic values, $\theta_L(\infty)$ and $t_L(\infty)$.
Per each $N$ a separate fitting procedure has been carried on, leading to different estimates for the asymptotic values.
In all cases, the quality of the fits is estimated by standard statistical analysis~\cite{NumericalRecipes}:
normalized $\chi$-square test ${\tilde \chi}^2 \equiv \frac{\chi^2}{D-f}$,
where $D - f$ is the difference between the number of data points, $D$, and the number of fit parameters, $f$.
The corresponding $\mathcal Q(D-f, \chi^2)$-values provide a quantitative indicator for the likelihood that $\chi^2$ should exceed the observed value, if the model were correct~\cite{NumericalRecipes}.
The results of all fits are reported together with the corresponding errors, ${\tilde \chi}^2$ and $\mathcal Q$ values (Table~\ref{tab:ThetaLTL-Fits}).
The reported value for $\Delta$ corresponds to the one giving the smallest ${\tilde \chi}^2$.
Our final values for $\theta_L$ and $t_L$ (Table~\ref{tab:ThetaLTL-Fits}, in boldface) are hence calculated by averaging the separate results for different $N$'s.
Error bars are given as $\sqrt{(\mbox{statistical error})^2 + (\mbox{systematic error})^2}$, where
the ``statistical error'' is the largest error bar between different $N$'s~\cite{MadrasJPhysA1992}
and
the ``systematic error'' is the spread between the single estimates.

Unfortunately, excluding the exceptions of $\theta_r$ for ideal trees (all $N$'s) and for $3d$ self-avoiding trees ($N=450$), the same strategy applied to the pair $(\theta_r, t_r)$ fails (Table~\ref{tab:ThetaRingTRing-Fits}).
This is probably related to the fact that finite size effects for $\theta_r$ and $t_r$ are, in general, smaller than for the previous exponents.
In those cases where the extrapolation procedure does not work, we have then simply taken averages of single estimates for each different $N$ and calculated error bars by the same combination of statistical and systematic errors as described above.
Final values for $\theta_r$ and $t_r$ (Table~\ref{tab:ThetaRingTRing-Fits}, in boldface) are then computed in the same way as for $\theta_L$ and $t_L$.

%%%
\section{Supplementary tables}\label{sec:ScalingExponents}
%%%

%
\begin{table*}
\begin{tabular}{ccccccccc}
& \multicolumn{2}{c}{$3d$ ideal trees} & \multicolumn{2}{c}{$3d$ self-avoiding trees} & \multicolumn{2}{c}{$2d$ melt of trees} & \multicolumn{2}{c}{$3d$ melt of trees} \\ 
\hline
\hline
\\
\\
\multicolumn{9}{c}{(a) $\theta_L$ and $t_L$ from best fits of numerical distributions, $p_N(L | \ell)$, to the RdC function} \\
\hline
$\ell$ & $\theta_L$ & $t_L$ & $\theta_L$ & $t_L$ & $\theta_L$ & $t_L$ & $\theta_L$ & $t_L$ \\
\hline
\\
\multicolumn{9}{c}{$N=450$}\\
\hline
$32$ & $1.250\pm0.041$ & $2.851\pm0.094$ & $1.013\pm0.034$ & $3.459\pm0.125$ & $1.221\pm0.041$ & $3.078\pm0.108$ & $1.223\pm0.040$ & $2.941\pm0.098$ \\
$64$ & $1.478\pm0.033$ & $2.418\pm0.053$ & $1.148\pm0.025$ & $2.878\pm0.063$ & $1.366\pm0.034$ & $2.639\pm0.066$ & $1.426\pm0.033$ & $2.483\pm0.056$ \\
$128$ & $1.632\pm0.027$ & $2.251\pm0.035$ & $1.211\pm0.021$ & $2.698\pm0.045$ & $1.406\pm0.026$ & $2.528\pm0.046$ & $1.555\pm0.027$ & $2.313\pm0.038$ \\
\hline
$\Delta$ & $0.569$ & $1.374$ & $1.104$ & $1.700$ & $1.862$ & $1.972$ & $0.655$ & $1.436$ \\
$\tilde{\chi}^2$ & $4\times10^{-7}$ & $2\times10^{-7}$ & $5\times10^{-9}$ & $8\times10^{-8}$ & $4\times10^{-8}$ & $5\times10^{-8}$ & $2\times10^{-7}$ & $7\times10^{-8}$ \\
$\mathcal{Q}$ & $0.9995$ & $0.9997$ & $0.9999$ & $0.9998$ & $0.9998$ & $0.9998$ & $0.9996$ & $0.9998$ \\
$\ell\rightarrow\infty$ & $1.95 \pm 0.06$ & $2.15 \pm 0.04$ & $1.27 \pm 0.03$ & $2.62 \pm 0.05$ & $1.42 \pm 0.03$ & $2.49 \pm 0.05$ & $1.78 \pm 0.05$ & $2.21 \pm 0.04$ \\
\hline
\\
\multicolumn{9}{c}{$N=900$}\\
\hline
$32$ & $1.259\pm0.041$ & $2.845\pm0.094$ & $1.026\pm0.032$ & $3.514\pm0.120$ & $1.248\pm0.041$ & $3.065\pm0.106$ & $1.231\pm0.040$ & $2.937\pm0.097$ \\
$64$ & $1.517\pm0.037$ & $2.370\pm0.055$ & $1.191\pm0.026$ & $2.852\pm0.063$ & $1.409\pm0.038$ & $2.616\pm0.070$ & $1.472\pm0.038$ & $2.432\pm0.061$ \\
$128$ & $1.711\pm0.033$ & $2.135\pm0.039$ & $1.292\pm0.025$ & $2.608\pm0.049$ & $1.523\pm0.034$ & $2.374\pm0.050$ & $1.637\pm0.033$ & $2.203\pm0.042$ \\
$256$ & $1.822\pm0.024$ & $2.053\pm0.025$ & $1.310\pm0.017$ & $2.503\pm0.031$ & $1.499\pm0.023$ & $2.402\pm0.036$ & $1.717\pm0.024$ & $2.137\pm0.028$ \\
\hline
$\Delta$ & $0.571$ & $1.197$ & $1.162$ & $1.366$ & $1.607$ & $ 1.791$ & $0.741$ & $1.352$ \\
$\tilde{\chi}^2$ & $5\times10^{-2}$ & $5\times10^{-2}$ & $2\times10^{-1}$ & $9\times10^{-3}$ & $7\times10^{-1}$ & $9\times10^{-1}$ & $6\times10^{-2}$ & $7\times10^{-2}$ \\
$\mathcal{Q}$ & $0.9509$ & $0.9477$ & $0.7924$ & $0.9906$ & $0.4884$ & $0.4121$ & $0.9462$ & $0.9349$ \\
$\ell\rightarrow\infty$ & $2.07 \pm 0.04$ & $1.98 \pm 0.03$ & $1.34 \pm 0.02$ & $2.44 \pm 0.03$ & $1.52 \pm 0.02$ & $2.37 \pm 0.03$ & $1.85 \pm 0.03$ & $2.08 \pm 0.03$ \\
\hline
\\
\multicolumn{9}{c}{$N=1800$}\\
\hline
$32$ & $1.265\pm0.041$ & $2.835\pm0.093$ & $1.048\pm0.034$ & $3.502\pm0.125$ & $ $ & $ $ & $ $ & $ $ \\
$64$ & $1.524\pm0.037$ & $ 2.351\pm0.054$ & $1.214\pm0.026$ & $2.885\pm0.062$ & $ $ & $ $ & $ $ & $ $ \\
$128$ & $1.782\pm0.037$ & $2.042\pm0.039$ & $1.344\pm0.025$ & $2.585\pm0.047$ & $ $ & $ $ & $ $ & $ $ \\
$256$ & $1.922\pm0.031$ & $1.942\pm0.029$ & $1.399\pm0.023$ & $2.473\pm0.039$ & $ $ & $ $ & $ $ & $ $ \\
$512$ & $1.963\pm0.021$ & $1.945\pm0.019$ & $1.359\pm0.017$ & $2.528\pm0.031$ & $ $ & $ $ & $ $ & $ $ \\
\hline
$\Delta$ & $0.651$ & $1.317$ & $1.247$ & $1.575$ \\
$\tilde{\chi}^2$ & $9\times10^{-1}$ & $1\times10^{0}$ & $2\times10^{0}$ & $1\times10^{0}$ \\
$\mathcal{Q}$ & $0.4397$ & $0.4122$ & $0.144$ & $0.320$ \\
$\ell\rightarrow\infty$ & $2.12 \pm 0.02$ & $1.91 \pm 0.02$ & $1.39 \pm 0.01$ & $2.48 \pm 0.02$ \\
\hline
\\
& {$\mathbf{2.05 \pm 0.09}$} & {$\mathbf{2.01 \pm 0.11}$} & {$\mathbf{1.33 \pm 0.06}$} & {$\mathbf{2.51 \pm 0.09}$} & {$\mathbf{1.47 \pm 0.06}$} & {$\mathbf{2.43 \pm 0.08}$} & {$\mathbf{1.82 \pm 0.06}$} & {$\mathbf{2.15 \pm 0.08}$} \\
\hline
\\
\\
\multicolumn{9}{c}{(b) Theoretical predictions $\theta_L = \frac{2}{\rho}-2$ and $t_L = \frac{1}{1-\rho}$} \\
\hline
& $\theta_L$ & $t_L$ & $\theta_L$ & $t_L$ & $\theta_L$ & $t_L$ & $\theta_L$ & $t_L$ \\
\hline
& $\mathbf{2.08\pm0.34}$ & $\mathbf{1.96\pm0.15}$ & $\mathbf{1.13\pm0.10}$ & $\mathbf{2.78\pm0.15}$ & $\mathbf{1.26\pm0.04}$ & $\mathbf{2.58\pm0.05}$ & $\mathbf{1.85\pm0.37}$ & $\mathbf{2.08\pm0.22}$ \\
\hline
\end{tabular}
\caption{
\label{tab:ThetaLTL-Fits}
Conformational statistics of tree contour distance $L$ at given ring contour distance $\ell$.
(a)
Effective exponents $\theta_L$ and $t_L$ obtained by best fits of numerical distributions, $p_N(L | \ell)$ (Fig.~\ref{fig:LtreesPDFs}), to the Redner-des Cloizeaux function, Eq.~(\ref{eq:q_RdC_L}),
for different $\ell$ and tree sizes $N$.
Final estimates and corresponding error bars 
are highlighted in boldface.
(b)
Estimates of exponents $\theta_L$ and $t_L$ according to the scaling relations Eqs.~(\ref{eq:ThetaLvsRho}) and~(\ref{eq:TLvsRho})
and using average values and error bars of exponents $\rho$ from our works~\cite{RosaEveraers-JPA2016,RosaEveraers-JCP2016}:
$\rho = 0.49\pm0.04$ ($3d$ ideal trees),
$\rho = 0.64\pm0.02$ ($3d$ self-avoiding trees),
$\rho = 0.613\pm0.007$ ($2d$ tree melt),
$\rho = 0.52\pm0.05$ ($3d$ tree melt).
}
\end{table*}
\begin{table*}
\begin{tabular}{ccccccccc}
\multicolumn{1}{c}{} & \multicolumn{2}{c}{$3d$ ideal trees} & \multicolumn{2}{c}{$3d$ self-avoiding trees} & \multicolumn{2}{c}{$2d$ melt of trees} & \multicolumn{2}{c}{$3d$ melt of trees} \\ 
\hline
\hline
\\
\\
\multicolumn{9}{c}{(a) $\theta_r$ and $t_r$ from best fits of numerical distributions, $p_N(\vec r | \ell)$, to the RdC function} \\
\hline
$\ell$ & $\theta_r$ & $t_r$ & $\theta_r$ & $t_r$ & $\theta_r$ & $t_r$ & $\theta_r$ & $t_r$ \\
\hline
\\
\multicolumn{9}{c}{$N=450$}\\
\hline
$32$ & $0.541\pm0.139$ & $1.251\pm0.046$ & $-0.106\pm0.017$ & $1.991\pm0.020$ & $0.400\pm0.015$ & $1.870\pm0.017$ & $0.245\pm0.034$ & $1.563\pm0.019$ \\
$64$ & $0.236\pm0.057$ & $1.321\pm0.025$ & $-0.136\pm0.008$ & $2.001\pm0.011$ & $0.420\pm0.009$ & $1.877\pm0.012$ & $0.265\pm0.010$ & $1.534\pm0.006$ \\
$128$ & $0.113\pm0.023$ & $1.373\pm0.012$ & $-0.159\pm0.006$ & $2.032\pm0.009$ & $0.396\pm0.005$ & $1.968\pm0.007$ & $0.278\pm0.005$ & $1.536\pm0.004$ \\
\hline
$\Delta$ & $1.309$ & -- & $0.435$ & -- & -- & -- & $0.622$ & -- \\
$\tilde{\chi}^2$ & $4\times10^{-8}$ & -- & $8\times10^{-9}$ & -- & -- & -- & $4\times10^{-9}$ & -- \\
$\mathcal{Q}$ & $0.9998$ & -- & $0.9999$ & -- & -- & -- & $0.9999$ & -- \\
$\ell\rightarrow\infty$ & $0.03\pm0.04$ & $1.32\pm0.07$ & $-0.22\pm0.02$ & $2.01\pm0.03$ & $0.41\pm0.02$ & $1.91\pm0.05$ & $0.30\pm0.02$ & $1.54\pm0.02$ \\
\hline
\\
\multicolumn{9}{c}{$N=900$}\\
\hline
$32$ & $0.464\pm0.131$ & $1.277\pm0.046$ & $-0.081\pm0.019$ & $1.962\pm0.021$ & $0.409\pm0.014$ & $1.866\pm0.015$ & $0.204\pm0.031$ & $1.595\pm0.018$ \\
$64$ & $0.208\pm0.056$ & $1.332\pm0.025$ & $-0.087\pm0.010$ & $1.935\pm0.013$ & $0.457\pm0.011$ & $1.821\pm0.013$ & $0.243\pm0.008$ & $1.550\pm0.005$ \\
$128$ & $0.150\pm0.024$ & $1.344\pm0.012$ & $-0.127\pm0.006$ & $1.979\pm0.008$ & $0.448\pm0.007$ & $1.879\pm0.009$ & $0.287\pm0.007$ & $1.525\pm0.005$ \\
$256$ & $0.077\pm0.010$ & $1.392\pm0.006$ & $-0.163\pm0.005$ & $2.077\pm0.008$ & $0.421\pm0.005$ & $1.974\pm0.007$ & $0.300\pm0.003$ & $1.524\pm0.002$ \\
\hline
$\Delta$ & $0.735$ & -- & -- & -- & -- & -- & $0.842$ & $1.590$ \\
$\tilde{\chi}^2$ & $4\times10^{-1}$ & -- & -- & -- & -- & -- & $7\times10^{-1}$ & $8\times10^{-1}$ \\
$\mathcal{Q}$ & $0.6962$ & -- & -- & -- & -- & -- & $0.5000$ & $0.4424$ \\
$\ell\rightarrow\infty$ & $-0.02\pm0.03$ & $1.34\pm0.06$ & $-0.11\pm0.04$ & $1.99\pm0.06$ & $0.43\pm0.02$ & $1.89\pm0.06$ & $0.324\pm0.005$ & $1.521 \pm 0.002$ \\
\hline
\\
\multicolumn{9}{c}{$N=1800$}\\
\hline
$32$ & $0.271\pm0.117$ & $1.361\pm0.046$ & $-0.089\pm0.013$ & $1.991\pm0.016$ &  &  & $ $ & $ $ \\
$64$ & $0.163\pm0.050$ & $1.354\pm0.023$ & $-0.079\pm0.008$ & $1.939\pm0.011$ &  &  & $ $ & $ $ \\
$128$ & $0.113\pm0.021$ & $1.365\pm0.011$ & $-0.093\pm0.006$ & $1.960\pm0.008$ &  &  & $ $ & $ $ \\
$256$ & $0.086\pm0.009$ & $1.379\pm0.006$ & $-0.085\pm0.005$ & $1.988\pm0.007$ &  &  & $ $ & $ $ \\
$512$ & $0.069\pm0.004$ & $1.393\pm0.003$ & $-0.110\pm0.003$ & $2.065\pm0.005$ &  &  & $ $ & $ $ \\
\hline
$\Delta$ & $0.827$ & -- & -- & -- &  &  & $ $ & $ $ \\
$\tilde{\chi}^2$ & $9\times10^{-3}$ & -- & -- & -- &  &  & $ $ & $ $ \\
$\mathcal{Q}$ & $0.9988$ & -- & -- & -- &  &  & $ $ & $ $ \\
$\ell\rightarrow\infty$ & $0.05\pm0.01$ & $1.37\pm0.05$ & $-0.09\pm0.02$ & $1.99\pm0.05$ &  &  & $ $ & $ $ \\
\hline
\\
\multicolumn{1}{c}{} & {$\mathbf{0.02\pm0.05}$} & {$\mathbf{1.34\pm0.07}$} & {$\mathbf{-0.14\pm0.07}$} & {$\mathbf{2.00\pm0.06}$} & {$\mathbf{0.42\pm0.02}$} & {$\mathbf{1.90\pm0.06}$} & {$\mathbf{0.31\pm0.05}$} & {$\mathbf{1.53\pm0.02}$} \\
\hline
\\
\\
\multicolumn{9}{c}{(b) Theoretical predictions $\theta_{r} = \min( \theta_{path} , \frac{2-\rho}{\nu} - d )$ and $t_r = \frac{1}{1-\nu}$} \\
\hline
& $\theta_r$ & $t_r$ & $\theta_r$ & $t_r$ & $\theta_r$ & $t_r$ & $\theta_r$ & $t_r$ \\
\hline
& $\mathbf{0}$ & $\mathbf{1.33\pm0.04}$ & $\mathbf{-0.17 \pm 0.28}$ & $\mathbf{1.92\pm0.15}$ & $\mathbf{0.63\pm0.04}$ & $\mathbf{1.92\pm0.07}$ & $\mathbf{0.28\pm0.01}$ & $\mathbf{1.47\pm0.04}$ \\
\hline
\end{tabular}
\caption{
\label{tab:ThetaRingTRing-Fits}
Conformational statistics of end-to-end spatial distances of ring sections of linear size $\ell$.
(a)
Effective exponents $\theta_r$ and $t_r$ obtained by best fits of numerical distributions, $p_N(\vec r | \ell)$ (Fig.~\ref{fig:IntDistsPDFs}), to the Redner-des Cloizeaux function, Eq.~(\ref{eq:q_RdC_r}),
for different $\ell$ and tree sizes $N$.
Final estimates and corresponding error bars 
are highlighted in boldface.
(b)
Estimates of exponents $\theta_r$ and $t_r$ according to the scaling relations Eqs.~(\ref{eq:tTree}) and~(\ref{eq:ThetaTree})
and using average values and error bars of exponents $\rho$ (summarized in the caption of Table~\ref{tab:ThetaLTL-Fits}), $\nu$ and $\theta_{path}$ from our works~\cite{RosaEveraers-JPA2016,RosaEveraers-JCP2016,RosaEveraers-PRE2017}:
$\nu = 0.25\pm0.02$ and $\theta_{path} = 0$ ($3d$ ideal trees),
$\nu = 0.48\pm0.04$ and $\theta_{path} = 1.07\pm0.08$ ($3d$ self-avoiding trees),
$\nu = 0.48\pm0.02$ and $\theta_{path} = 0.63\pm0.04$ ($2d$ tree melt),
$\nu = 0.32\pm0.02$ and $\theta_{path} = 0.28\pm0.01$ ($3d$ tree melt),
}
\end{table*}
\end{document}